\documentclass[12pt]{article}
\usepackage{epsfig}

\def\beq{\begin{equation}}
\def\eeq#1{\label{#1}\end{equation}}
\def\eeqn{\end{equation}}

\def\beqa{\begin{eqnarray}}
\def\eeqa#1{\label{#1}\end{eqnarray}}
\def\eeqan{\end{eqnarray}}

\let\bar=\overbar

\def\Dslash{\not{\hbox{\kern-4pt $D$}}}
\def\dslash{\not{\hbox{\kern-2pt $\del$}}}

\def\msb{{\bar{\ssstyle M \kern -1pt S}}}

\def\Title#1{\begin{center} {\Large {\bf #1} } \end{center}}

\begin{document}

\Title{Recent CMS Results on Forward and Small-$x$ QCD Physics}

\bigskip\bigskip

\addtocontents{toc}{{\it B. Roland}}
\label{RolandStart}

\begin{raggedright}  

{\it Roland Beno\^it\index{Roland, B.} } on behalf of the CMS Collaboration\\
{\it Physics Department\\
University of Antwerp \\
B-2020 Antwerpen Belgium}
\bigskip\bigskip

\end{raggedright}

\vspace*{-0.4cm}
\section{Introduction}

Recent CMS results on Forward and Small-$x$ QCD Physics are presented. Those include the measurement of the Underlying Event activity
and the study of jet production at large rapidity separation. 

\section{Measurement of the UE activity}

\begin{figure}[htb]
\begin{center}
\hspace*{-2.4cm}
\epsfig{file=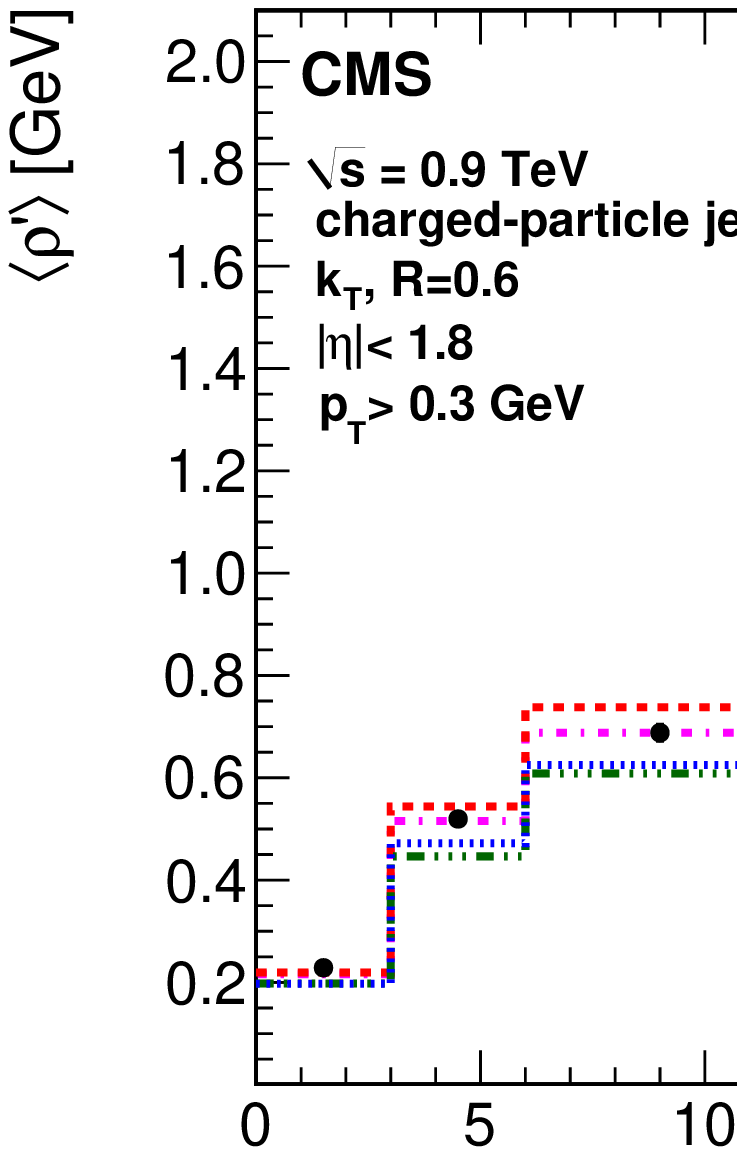,height=1.7in} 
\hspace*{2.5cm}
\epsfig{file=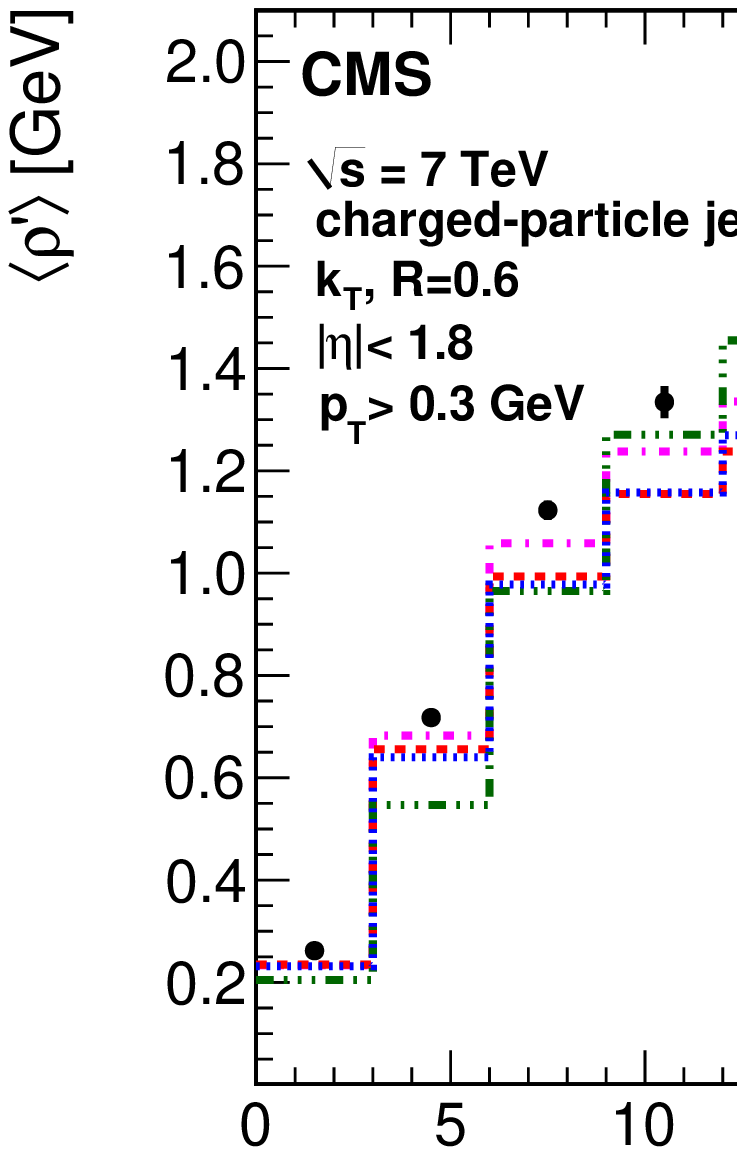,height=1.7in}
\hspace*{2.5cm}
\epsfig{file=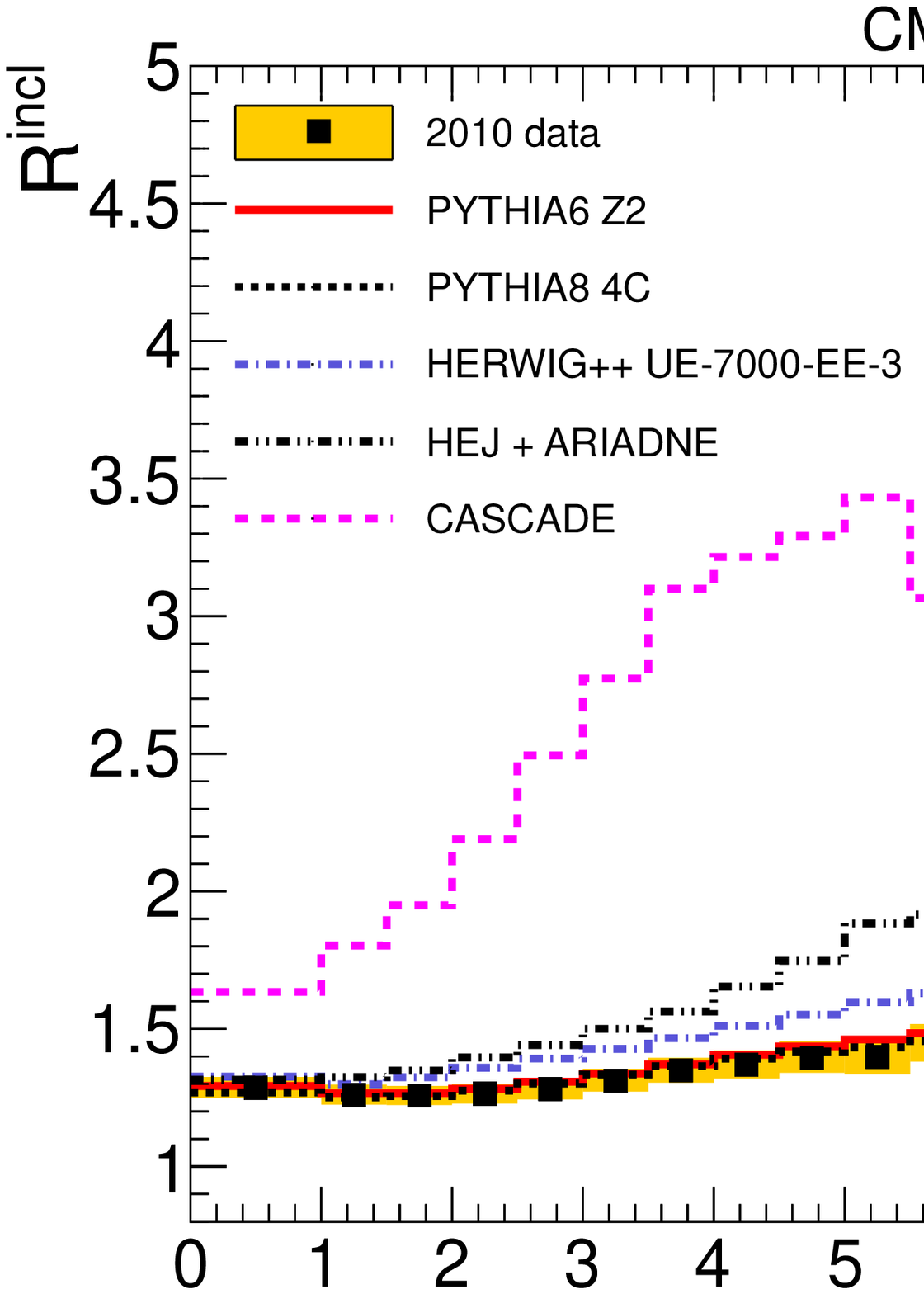,height=1.75in}
\caption{Mean values of the $\rho^\prime$ distributions versus leading charged particle jet $p_T$ at $\sqrt{s} = 900$ GeV (left) and 7 TeV (middle).
Inclusive to exclusive dijet cross sections ratio (right) as function of the rapidity separation between jets.}
\label{UE_Area_KF}
\end{center}
\end{figure}

The Underlying Event (UE) in a $p \, p$ collision is everything except the hard scattering. It includes the initial state radiation (ISR), 
the final state radiation (FSR), the multiple partonic interactions (MPI) and the beam remnants. A good understanding of the UE activity is important 
in order to extend our knowledge of Quantum Chromodynamics (QCD) towards the soft limit, for precision measurements of Standard Model processes and searches 
for new physics. Since the UE is produced by soft or semi-hard interactions, its dynamics can not be fully described by perturbative QCD. As a consequence 
phenomenological models have to be used whose parameters have to be tuned to data. \\ \newline
The measurement of the UE requires a physically motivated separation between the soft and hard components of the collision. In the central region, 
where there is no separation in pseudorapidity $\eta$ between the hard scattering and the UE, the traditional approach is to divide the phase-space in different 
$\varphi$ regions wrt. the leading $p_T$ object. The UE activity is then investigated through the measurement of the energy and particle densities 
in the so-called towards, transverse and away regions, respectively defined by $|\Delta \varphi| < 60^\circ$,  $60^\circ < |\Delta \varphi| < 120^\circ$ 
and $|\Delta \varphi| > 120^\circ$. This approach is followed in the study of the UE activity in the Drell-Yan (DY) process 
$q \bar{q} \to \gamma^*/Z^{(*)} \to \mu^+ \mu^-$~\cite{Chatrchyan:2012tb} which offers a clean separation between the hard scattering and the soft component, 
is experimentally clean and theoretically well understood. There is furthermore no QCD FSR and limited brehmsstrahlung from the muons, 
which are required to have $p_T > 20$ GeV/c and $|\eta| < 2.4$. The UE activity is determined from the primary charged particles with  
$p_T > 0.5$ GeV/c and $|\eta| < 2$ and is measured as a function of the dimuon invariant mass $M_{\mu \mu}$ and transverse momentum $p_T^{\mu \mu}$. 
\begin{figure}[htb]
\begin{center}
\epsfig{file=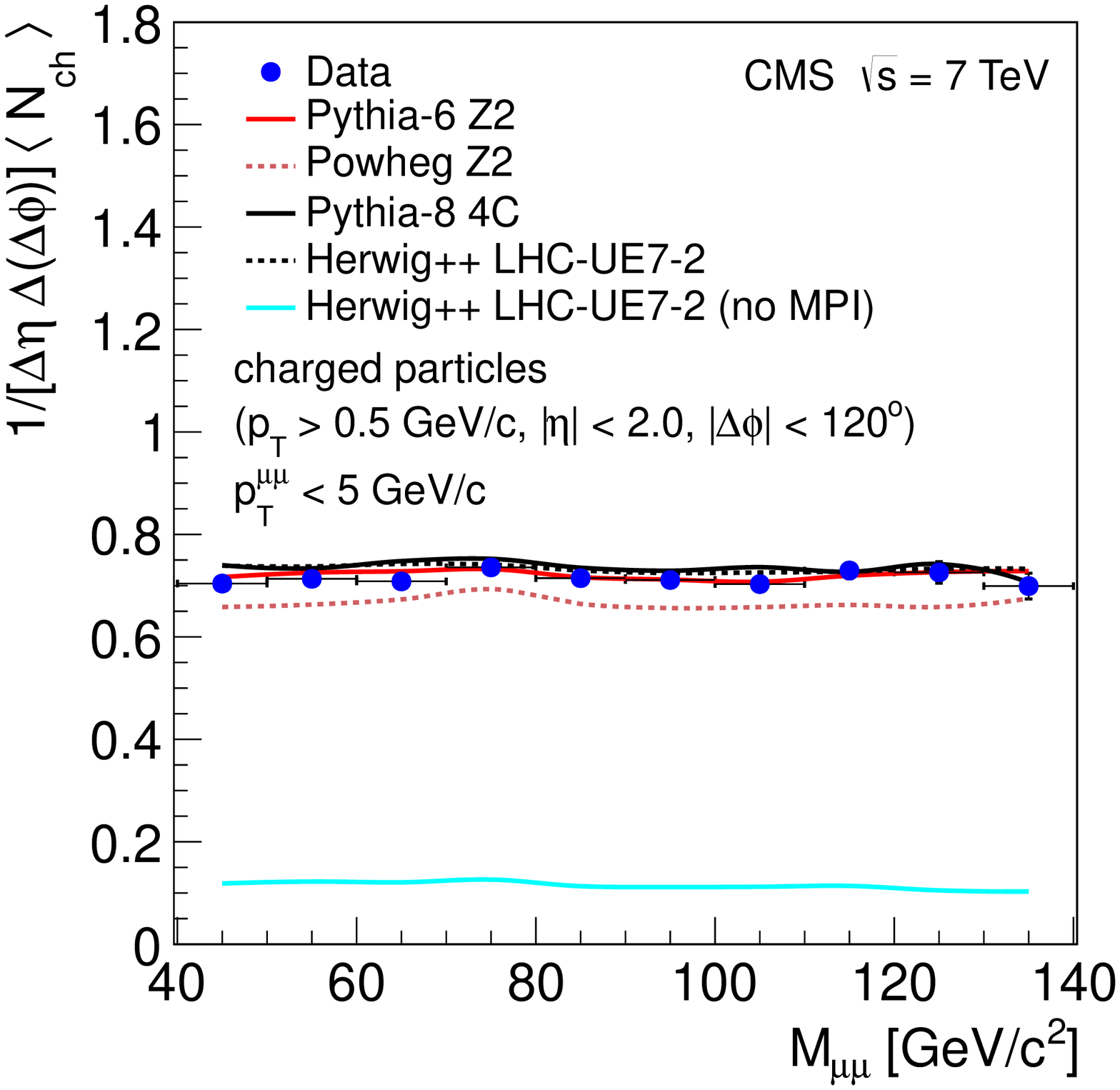,height=1.9in,clip=}  
\epsfig{file=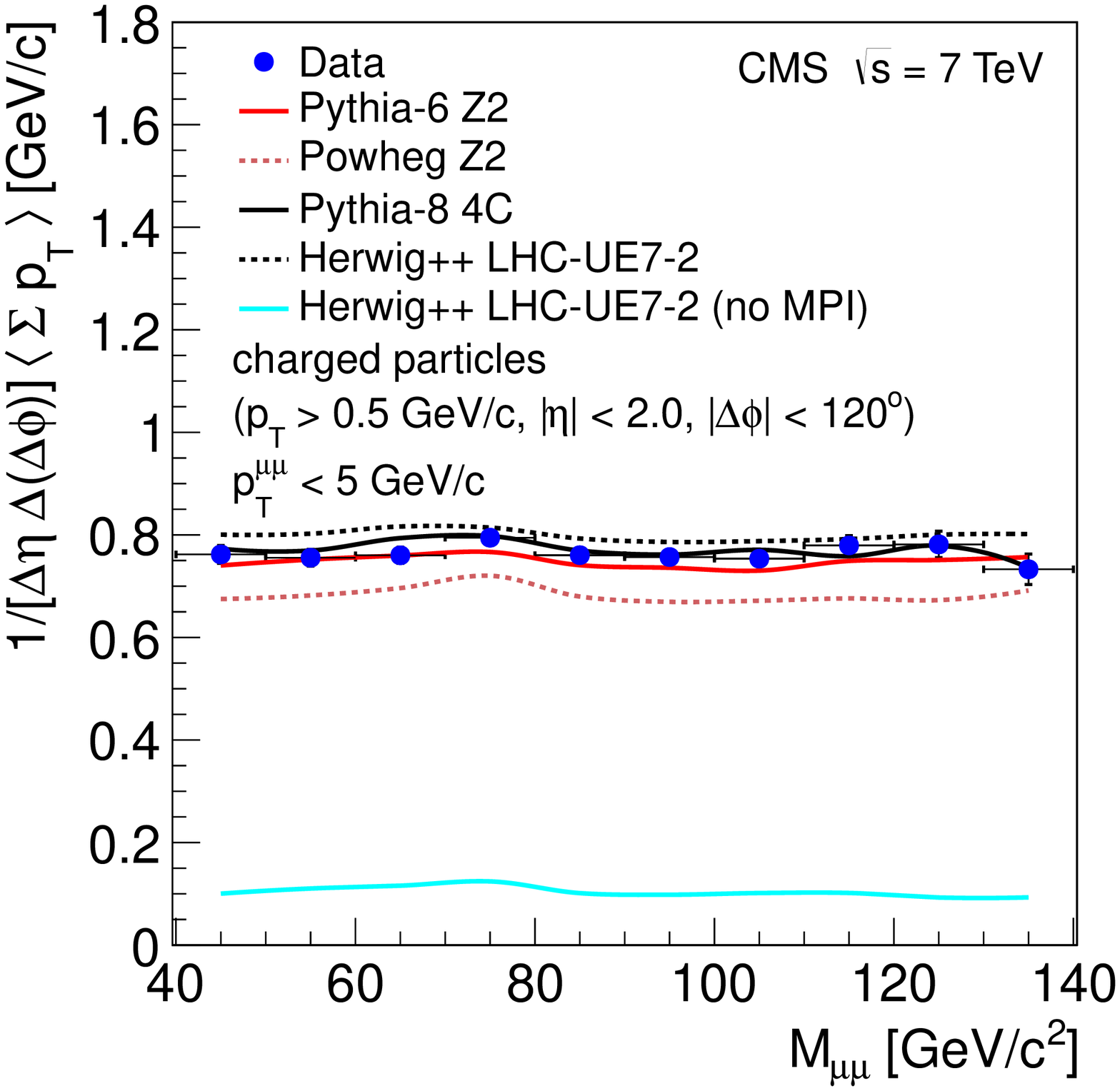,height=1.9in,clip=} 
\epsfig{file=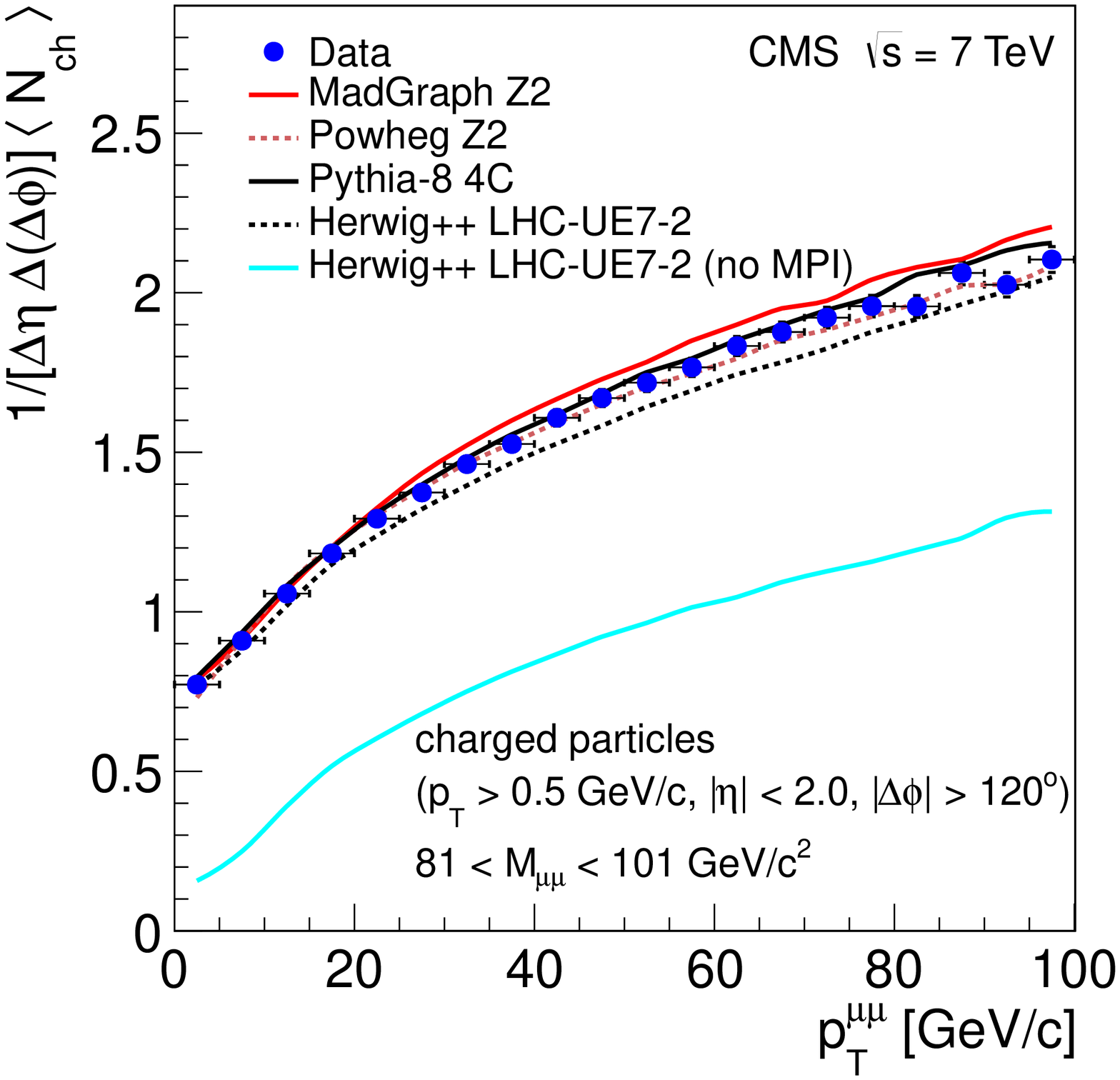,height=1.9in,clip=}           
\vspace*{-0.4cm}
\caption{Particle density (left) and energy density (middle) as a function of $M_{\mu \mu}$ in the towards and transverse regions 
$|\Delta \varphi| < 120^\circ$, with $p_T^{\mu \mu} < 5$ GeV/c. Particle density as a function of $p_T^{\mu \mu}$ in the away region (right), 
with $81 < M_{\mu \mu} < 101$ GeV/$c^2$.}
\label{UE_DY_density}
\end{center}
\vspace*{-0.4cm}
\end{figure}
Figure~\ref{UE_DY_density} shows the particle density (left) and energy density (middle) at $\sqrt{s} = 7$~TeV as a function of $M_{\mu \mu}$ 
in the towards and transverse regions. The condition $p_T^{\mu \mu} < 5$ GeV/c is applied to limit the ISR contribution and the UE is therefore 
dominated by the MPI, which account for approximately $80 \%$ of the activity. The presence of a hard scale given by the $Z$ boson virtuality 
centers the collision and saturates the MPI. Figure~\ref{UE_DY_density} also shows the particle density as function of $p_T^{\mu \mu}$ in the away region (right), 
in the range $81 < M_{\mu \mu} < 101$ GeV/$c^2$. Since the MPI is saturated at this energy scale, the $p_T^{\mu \mu}$ dependence is sensitive to the ISR contribution. 
The density rises sharply with $p_T^{\mu \mu}$ in that region, which is mainly sensitive to the hardest ISR. The slope of the Monte Carlo (MC) prediction is 
unaffected by the presence of the saturated MPI, showing that the slope is mainly due to the ISR contribution. \\ \newline
An alternative approach to study the UE activity at central rapidity, for which no subdivision in $\varphi$ is needed, is to look at the jet $p_T$ 
per jet area in the $(\eta,\varphi)$ plane~\cite{2012tt}. The median $\rho^\prime$ of the distribution is a measure of the soft activity in each event 
and is robust to outliers, in particular to the leading jets.  The jets are reconstructed in the region $|\eta| < 1.8$ with the $k_T$ algorithm 
with a radius R = 0.6. Figure~\ref{UE_Area_KF} shows the mean values of the $\rho^\prime$ distributions as function of leading jet $p_T$ at $\sqrt{s} = 900$~GeV 
(left) and 7~TeV (middle). The typical UE behaviour manifests itself by a steep rise at small event scale, followed by a saturation of the activity 
at higher leading jet $p_T$. The PYTHIA6 Z1 tune gives the best description of the data, while it is difficult for the different 
tunes to reproduce the UE evolution with $\sqrt{s}$. \\ \newpage
\vspace*{-0.1cm}
\vspace*{-0.8cm}
\begin{figure}[htb]
\begin{center}
\epsfig{file=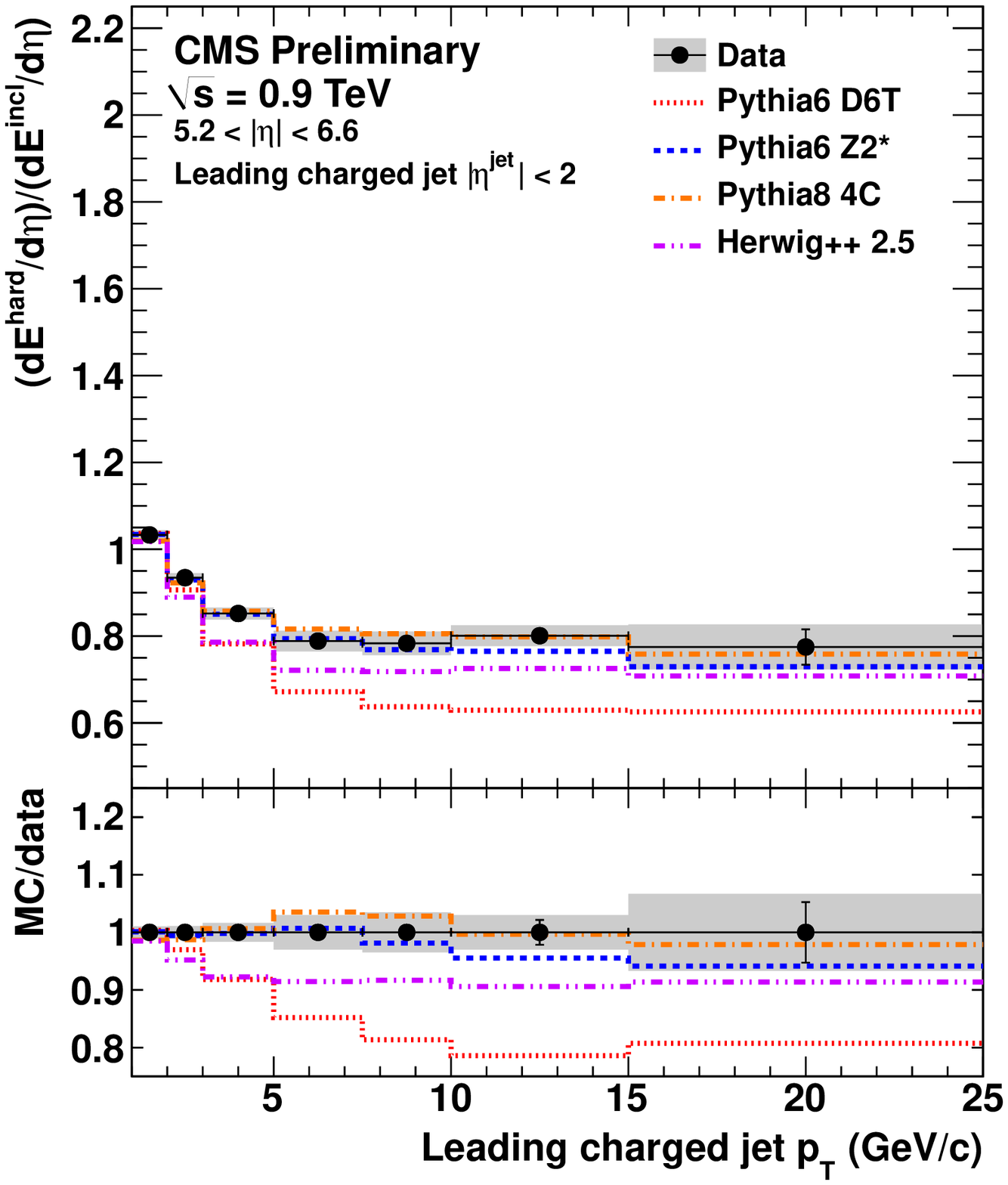,height=2.1in,clip=}         
\hspace*{0.5cm}
\epsfig{file=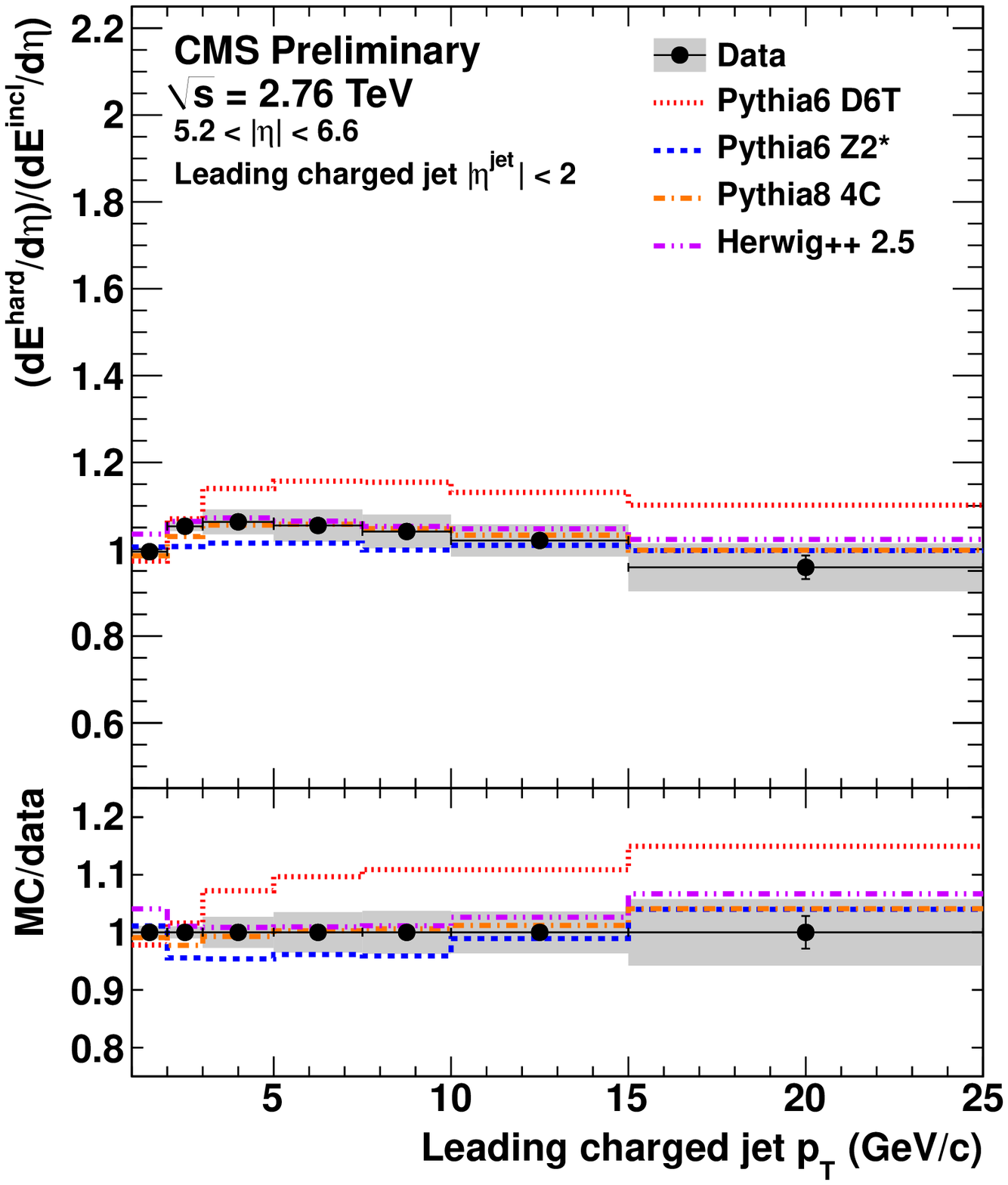,height=2.1in,clip=}        
\hspace*{0.5cm}
\epsfig{file=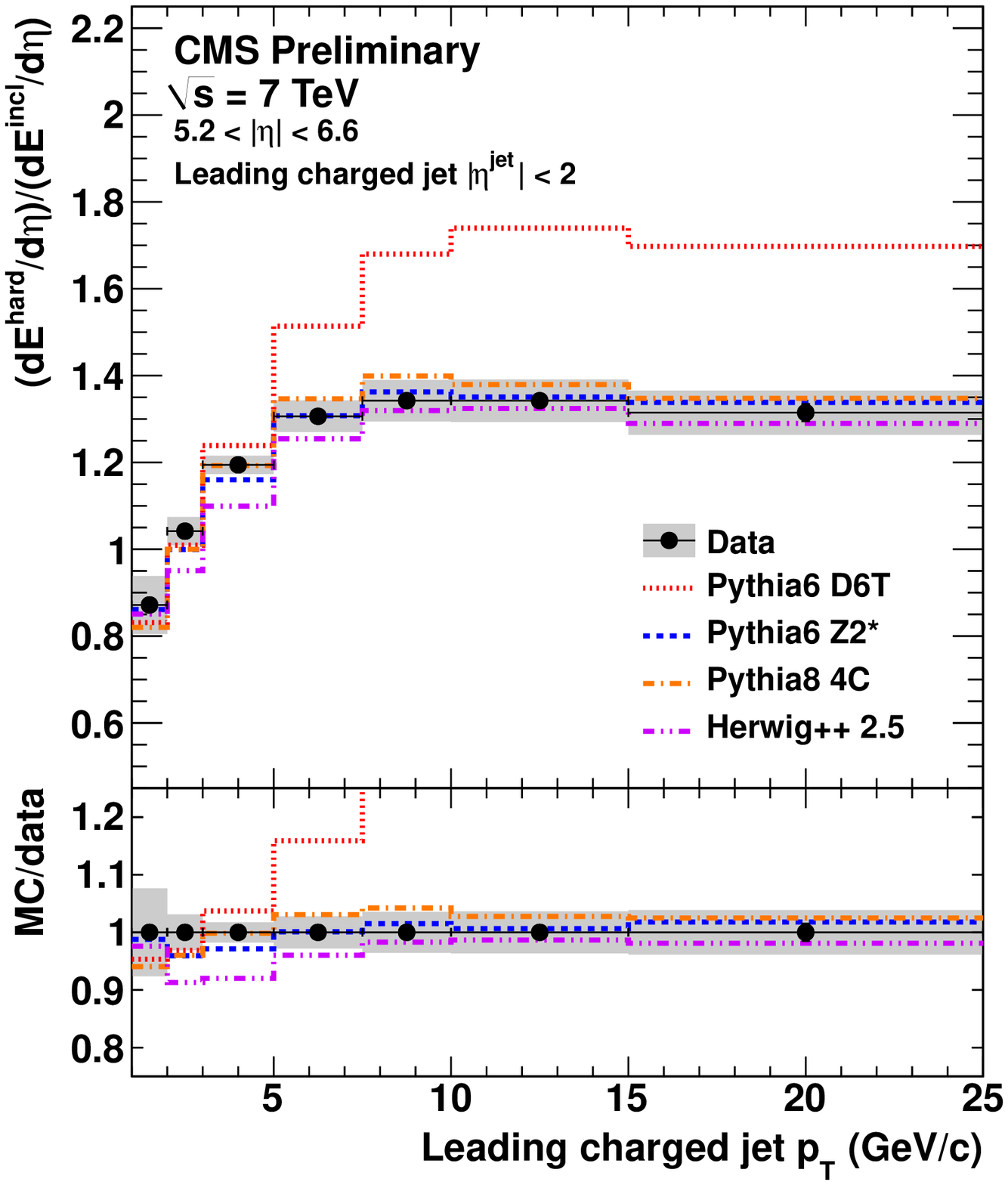,height=2.1in,clip=}           
\vspace*{-0.2cm}
\caption{Ratio of the energy density in the hard and inclusive samples as function of leading jet $p_T$ at $\sqrt{s} = 900$ GeV (left), 
2.76 TeV (middle) and 7 TeV (right).}
\label{UE_Castor}
\end{center}
\vspace*{-0.1cm}
\end{figure}
\vspace*{-0.4cm}
In the forward region, the hard scattering and the UE are separated by a large pseudorapidity interval $\Delta \eta$ 
which makes the phase-space subdivision in $\varphi$ unnecessary and offers the possibility to study the $\varphi$ dependence of the UE activity. 
In this approach, the UE activity is measured in the range $-6.6 < \eta < -5.2$ by comparing the energy density of two classes of events~\cite{UECastor}.
Inclusive events, where the energy density is not much affected by MPI are compared to events with a hard scale, in which MPI strongly affect the energy density.
The hard scale is defined by the $p_T$ of the central leading charged particle jet with $p_T > 1$ GeV/c and $|\eta| < 2$. The energy density in the hard sample
is normalized to the inclusive one, making the results independent of calibration and minimizing the systematic uncertainties.
Figure~\ref{UE_Castor} shows the ratio of the energy density in the hard and inclusive samples as function of leading charged particle jet $p_T$ 
at $\sqrt{s} = 900$ GeV (left), 2.76~TeV (middle) and 7~TeV (right). At 7 TeV, the typical UE behaviour shows itself by a fast rise at low $p_T$ followed 
by a plateau for $p_T > 8$ GeV. At 900 GeV, the energy density in the inclusive sample is bigger than the hard one and the ratio is smaller than unity.
At this energy, the proton remnant fragments in the $\eta$ range of the measurement, and the increase of the UE activity in the central region with 
increasing $p_T$ depletes the energy in the hard sample. At 2.76 TeV one observes a much reduced increase of the energy ratio and the activities in both
samples are close to each other. The data are well described by PYTHIA6 tunes fitted to LHC data, while the pre-LHC tune PYTHIA6 D6T predicts too much MPI.

\section{Jet production at large rapidity separation}

Dijet production is studied as function of the rapidity separation $\Delta y$ between the jets\cite{Chatrchyan:2012pb}, for jets with $p_T > 35$ GeV and $|y| < 4.7$. 
Two classes of events are considered: inclusive events, with at least one pair of jets, and exclusive events, with exactly one pair of jets. The ratio $R_{incl}$ 
of the cross section of all pairwise combinations of jets from the inclusive sample to the exclusive dijet cross section is presented on figure \ref{UE_Area_KF}
(right) as a function of the rapidity difference between the jets $|\Delta y|$. This observable is expected to be sensitive to effects beyond collinear factorization, 
as the radiation probability increases with $|\Delta y|$ rather than with $p_T$.
The predictions of the MC event generators PYTHIA6 and PYTHIA8 agree with the measurements, while the HERWIG++ predictions exhibit 
a more pronounced rise with $|\Delta y|$. The BFKL-motivated generators CASCADE 
and HEJ+ARIADNE predict for these ratios a significantly stronger rise than observed.


\vspace*{-0.3cm}
\begin{thebibliography}{99}

\bibitem{Chatrchyan:2012tb}
S.~Chatrchyan {\it et al.}  [CMS Collaboration],
Measurement of the underlying event in the Drell-Yan process in proton-proton collisions at $\sqrt{s}$ = 7 TeV,
arXiv:1204.1411 [hep-ex].
 
\bibitem{2012tt}
S.~Chatrchyan {\it et al.}  [CMS Collaboration],
Measurement of the underlying event activity in $p p$ collisions at $\sqrt{s} = 0.9$ and 7 TeV with the novel jet-area/median approach,
arXiv:1207.2392 [hep-ex].


\bibitem{UECastor}
Study of the underlying event at forward rapidity in $p p$ collisions at the LHC, 
CMS PAS FWD-11-003.

\bibitem{Chatrchyan:2012pb}
S.~Chatrchyan {\it et al.}  [CMS Collaboration],
Ratios of dijet production cross sections as a function of the absolute difference in rapidity between jets in proton-proton collisions at $\sqrt{s}$ = 7 TeV,
arXiv:1204.0696 [hep-ex].





\end{thebibliography}
\end{document}